\documentclass[10pt,sigconf]{acmart}

\usepackage[english]{babel}
\usepackage{blindtext}

\usepackage{import}
\usepackage{graphicx}
\graphicspath{ {images/} }
\usepackage{listings}
\usepackage{caption}
\usepackage{subcaption}
\usepackage{tabularx}

\renewcommand\footnotetextcopyrightpermission[1]{} 
\setcopyright{none}

\acmDOI{ }

\acmISBN{ }

\acmConference[]{}
\acmYear{ }
\copyrightyear{ }

\acmPrice{}

\begin{document}
\title{CageCoach: Sharing-Oriented Redaction-Capable Distributed Cryptographic File System}


\author{Jason Carpenter}
\email{CARPE415@umn.edu}
\affiliation{%
  \institution{University of Minnesota}
  \city{Minneapolis}
  \state{MN}
}

\author{Zhi-Li Zhang}
\email{zhzhang@cs.umn.edu}
\affiliation{%
  \institution{University of Minnesota}
  \city{Minneapolis}
  \state{MN}
  }

\renewcommand{\shortauthors}{}

\begin{abstract}
The modern data economy is built on sharing data. However, sharing data can be an expensive and risky endeavour. Existing sharing systems like Distributed File Systems provide full read, write, and execute Role-based Access Control (RBAC) for sharing data, but can be expensive and difficult to scale. Likewise such systems operate on a binary access model for their data, either a user can read all the data or read none of the data. This approach is not necessary for a more read-only oriented data landscape, and one where data contains many dimensions that represent a risk if overshared. In order to encourage users to share data and smooth out the process of accessing such data a new approach is needed. This new approach must simplify the RBAC of older DFS approaches to something more read-only and something that integrates redaction for user protections. 

To accomplish this we present CageCoach, a simple sharing-oriented Distributed Cryptographic File System (DCFS). CageCoach leverages the simplicity and speed of basic HTTP, linked data concepts, and automatic redaction systems to facilitate safe and easy sharing of user data. The implementation of CageCoach is available at 

\url{https://github.com/Jcarpenter-0/CageCoach}.
\end{abstract}

\maketitle


\section{Introduction}
User-generated data drives the modern world. Everything from Uber driver rides and Google search queries to video game experiences and Amazon purchase patterns feed user data back into these systems to provide insights for improvement. Additionally, users sharing their data as part of crowd sourcing solutions has proven key to reverse engineering gig working applications such as Uber\cite{reverse-engineer-uber-1,reverse-engineer-uber-lyft-1,data-platform-2-gridwise}, Lyft\cite{reverse-engineer-uber-lyft-1,data-platform-2-gridwise}, and Shipt\cite{reverse-engineering-shipt-1,data-platform-2-gridwise}. Further these efforts help solve civic and national needs such as with Atlanta's Data Dashboard\cite{community-data-atlanta}, Minneapolis's Opendata program\cite{collective-data-minneapolis}, or the United State's Citizen Science initiative\cite{collective-data-citizen-science}.

However, users providing their data to these initiatives often comes with a level of risk and a loss of control over the data they provide. Once a user has handed over information the safety considerations, redaction approaches, and management decisions are out of their control. Further, should any shared user data become dangerous to a user, the user has no more sway to alleviate this risk other than ask the current data holder to act, a practice often fruitless.

In order to further encourage users to share their data, a new sharing oriented data hosting system is required. Such a platform must be simple to implement, easy to request data from, but still provide some assurances of privacy and safety for users involved. Crucially it should remain in the user's control, and not be subject to control by others even those hosting data such as on public hosting systems. The privacy capability must be granular not just in who can access data but what specific data is accessible. For example, for some users, sharing their full name to everyone who asks is unreasonable. Thus they should be able to share with some a partial redaction of their name. Existing works such as Distributed File Systems (DFS) are promising, but require extensive implementation, Role-based Access Control (RBAC) enforcement, and do not implement granular redaction. Other platforms like Google Drive, Dropbox, and Kaggle are great for sharing bulk data but also do not provide granular redaction and require trusting of the platform holders to not share otherwise redacted user data.

In this work, we introduce CageCoach a sharing oriented distributed cryptographic file system. CageCoach's notable features are:

\noindent{\textbf{$\bullet$ Simple Trustless DCFS built over HTTP GET/POST}}

\noindent{\textbf{$\bullet$ Customizable RBAC and Datatype Granular Redaction Pipeline}}

\noindent{\textbf{$\bullet$ Easier sharing with Decentralized data access and centralized user control}}

CageCoach streamlines the older RBAC based models of DFSs and decentralizes the data hosting approaches of platforms making for an overall simpler means of sharing data with others while retaining granular privacy control for users. This system is leverages simple HTTP GET/POST operations to interact with symmetrically encrypted files hosted on any HTTP platform to achieve decentralized hosting. These files point back to their owners, represented by a controlling server, that can facilitate redacted data access for a data requester providing user control of data access. Finally, the user's controlling server applies user defined redaction operations from a suite of modules CageCoach provides to reduce sensitive data leakage.

CageCoach's code can be found at 

\url{https://github.com/Jcarpenter-0/CageCoach}.

\section{Related Work}
Distributed File Systems (DFS) and cryptographic file systems (DCFS) have been around for a long time with some works as early as 1993\cite{crypto-file-system-1} and as recent as 2020\cite{crypto-distributed-file-upss}. These are mature fields with well-developed and commercial products we see every day, such as Dropbox, GoogleDrive, Hadoop, Ceph, and others\cite{dfs-3-survey}. Despite this, the changing data landscape and changing usage behaviors with data invite a new perspective of existing systems to better fit them for a new era. The work must relevant in the current data landscape, data redaction, is an old field but with a renewed interest in the face of big data breaches, data privacy concerns, and machine learning for data protection. In this section we outline these two related areas and contrast them with our proposed system.

\subsection{Distributed \& Cryptographic File Systems}
Distributed File Systems (DFS) are systems for maintaining coherent file management across desperate hosting devices. Examples include standard file hosting such as Google Drive, Dropbox, and InRupt's Solid\cite{dfs-2-solid}. Such systems have a long history and continued relevance in the modern era. DFS also manifest as cloud storage systems, albeit with looser file system format adherence to mesh with the more diverse Internet access environment. Extending DFSs into privacy and security oriented spaces yields the Distributed Cryptographic File System (DCFS) domain. Works such as UPSS\cite{crypto-distributed-file-upss} focus on creating a sharing-oriented and protective DFS with full RBAC and mutable verifiable histories of each file involved as a check against malicious behavior. Further other works such as \cite{crypto-dfs-block-chain-oriented, crypto-distributed-3} aim to utilize the blockchain to achieve the same RBAC with a more decentralized approach. Finally, other approaches aim to refine key management in encryption for DFS\cite{crypto-distributed-4}.These systems while powerful, rely on relatively expensive RBAC and infrastructure or require significant trust for the platform holders. In the former case, simplifying the RBAC with the mostly read-only reality of user data can lower RBAC complexity significantly. In the latter case, hosting infrastructure is still necessary, but one must create a trustless environment in order to retain control of one's data even on such hosting platforms.

Our work focuses on streamlining data sharing by creating a middlepoint between strong, rigid, and RBAC focused approaches such as DCFSs and trust-oriented data platforms and services like Uber, Kaggle, and Gridwise.

\subsection{Data Redaction}
Data redaction is not a new field, but has gained vigor in the last decade or so as the data economy has shaped. Redaction provides the means for which sensitive data can be made less sensitive and thus less dangerous in the event of leaks, breaches, or theft. Likewise, redaction has its place in academic publications when such publications may contain in themselves dangerous or sensitive information\cite{redaction-in-academia}. Many existing tools provide a user the quick means of redacting a document such as \cite{redaction-tool-2} and \cite{redaction-tool-3}. A handful of commercial products, such as \cite{redaction-tool-1-doma}, \cite{redaction-tool-4}, and \cite{redaction-tool-5}, apply machine learning to identify and remove automatically sensitive data. Finally, other work such as \cite{transparent-redaction} highlight an interesting scenario where redaction itself must be transparent enough such that the redaction doesn't mislead the information. These systems as implemented are not part of a sharing pipeline and are applied ad-hoc to data. A system such as the one outlined by UPSS\cite{crypto-distributed-file-upss}, envisions such technologies are part of a pipeline of data requests but did not implement or specify beyond such designs.

 Our work applies the concepts behind these redaction systems, but crucially, as part of a standard granular access pipeline and not as a one-off and static redaction. This in effect realizes some aspects of the UPSS\cite{crypto-distributed-file-upss} pipeline, but without the more complex full RBAC suite.

\section{Problem And Design Goals}
 In order to build a system that encourages users to share their data two primary problems and design considerations must be achieved: Simplification of access control for accessing and requesting data and automatic policy informed data redaction. With these two aspects a sharing-oriented DFS will lower the cost of sharing and accessing data and provide a wide net of protections for users who choose to share.

\subsection{Simplify Access Control For Data}
Existing DFS systems utilize a full suite of RBAC functionality to provide read, write, and execute functionality for shared files. These provisions while useful, require significant infrastructure such as certificates and user profiles registered within the computational structure of the data host. This full suite of RBAC is necessary if the group of users intended to read, write, and/or execute the shared data, but costly if sharing (read only) is the intention. By removing the write and execute assumptions of RBAC we can in turn simplify the operating infrastructure required for accessing data and making sharing a lower cost effort. This lower cost is necessary for encouraging users to share their data, as it will be easier to host for consumption, and for consumers of data as it will be easier to access.

\subsection{Provide Integrated Automatic User Data Redaction}
Regardless of ease of access, users must be given some assurances of safety, privacy, and proper use for their data. Traditional RBAC focuses on binary access models for data, either a user can read all the data or none of the data in a typically hosted file. This approach is not adequate for data items that contain core sensitive fields. For example, a typical sales receipt is useful for inventory systems and market trending services, as they provide insights into purchases and sales trends, however, these same receipts may contain the purchaser's name, credit card information, and/or address and location. Such fields are not important for the overall trend, but present a security risk for the user. In a binary RBAC model, such fields would available if the receipt is available. A more granular approach to access is needed. Such an approach is outlined but not realized or specified by UPSS\cite{crypto-distributed-file-upss}. Such an approach would require that when a user's data is requested by another, a trusted middle system acquires the raw full set of data, and then redacts and removes information that is included in the data but not allowed for that particular user. For example, removing the name, address, and credit fields from the sales receipt scenario. This approach is required to provide granular and safer exposure of user's data for general consumption. Further, this process can be handled by user-defined policy thus providing guidelines for any user data added in the future thus lowering sharing costs further.

\section{CageCoach System}
\begin{figure}
    \centering
    \includegraphics[width=0.48\textwidth, height=0.31\textwidth]{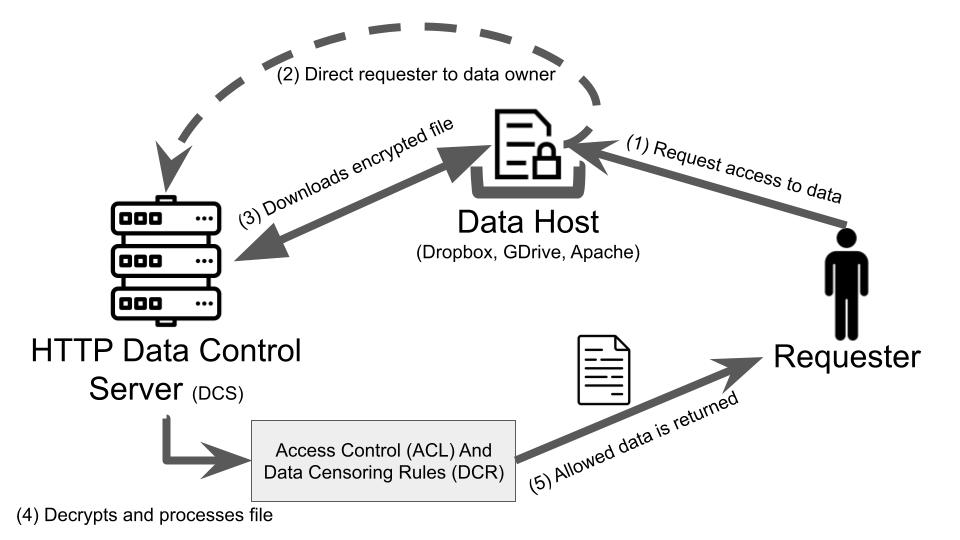}
    \caption{CageCoach System, providing a streamlined means for requestors to ask for data and receive useful but protected data.}
    \label{fig:system-overview}
\end{figure}

We realize the goals of a sharing-oriented DFS with our system CageCoach. CageCoach simplifies the RBAC and infrastructure of existing DFSs and integrates redaction technologies into a data request pipeline. All of this together creates a simple and easy means for users to safely and easily share their data. CageCoach is organized around several concepts and a flow, outlined in fig. \ref{fig:system-overview}. Requesters, who request user data. Data hosts, which host encrypted data files and some attached meta data files. Finally, a Data Control Server (DCS) which manages the owner's data, processes requests made by requesters, and redacts outgoing sensitive data. CageCoach's operational use-case is:

\begin{enumerate}
    \item A owner uploads some data (video, text, audio, etc) to a hosting system after encrypting and creating a meta file for the data.
    \item A requester sees this data and examines the meta file (using GET for example) for information as to where the owner's DCS operates.
    \item The requester sends a POST request to the owner's DCS server, asking to view the original data item.
    \item The DCS receives this request, verifies the requester's identity through asymmetric key phrase decryption, and then uses GET to retrieve the remotely hosted encrypted data file.
    \item The DCS decrypts the file with its own internal symmetric key and then applies a series of redaction operations on the data.
    \item The DCS forwards the remaining unredacted data to the requester, completing the request and preventing unnecessary or forbidden data from leaving encrypted/controlled space.
\end{enumerate}

The details for how the RBAC is simplified and how the redaction is integrated is detailed in the following sections.

\subsection{Simplifying RBAC Using HTTP And Read-Only Assumptions}
CageCoach simplifies the primary RBAC and infrastructure of other DFSs by assuming that user data need only be read, not written too or executed collaboratively. Additionally, unlike UPSS\cite{crypto-distributed-file-upss}, since there is no write permissions data versions are no longer necessary thus can relax the assumption UPSS makes for needing a transparent modifications tree. With this simplification in mind, CageCoach utilizes the most common means of read-only operation on the Internet: HTTP GET. This means that user data can be hosted on any system that facilitates HTTP GET, such as open source systems like Apache2. The data that gets hosted is the user's encrypted file and a plain text meta data file. Using some concepts of linked data, the meta data file points to the owner's DCS to actually facilitate the request for data among other fields. The total definition for this meta data file is:

\noindent{\textbf{$\bullet$ owner-url}}: URL indicating where the owner's DCS is. The place where any request will be processed.

\noindent{\textbf{$\bullet$ meta-data}}: User filled info tags about the data, such as what format it is, overall context. All of this information is optional.

\noindent{\textbf{$\bullet$ description}}: A more textual description of the data, optional if an owner wishes to provide more than just tags of information.

\noindent{\textbf{$\bullet$ data-url}}: The URL indicating where the data this meta file belongs to is. This is important for providing some backup if the meta file is moved elsewhere or if it must live elsewhere in hosting.

\noindent{\textbf{$\bullet$ data-hash-sha1}}: A sha1 of the encrypted file to provide a minimal check for any requester that wishes to double check the file they are asking about.

Despite our overall read-only approach, some computational efforts are still required. Namely the decryption of the requested file and the granular redaction of information within this file. The purpose of redirecting the requester from the data host is to provide a centralized response by the owner and the computational space for redaction policies. The requester will send an HTTP POST request to the DCS indicated by the owner-url and receive a decrypted and redacted data file. The DCS's process is implemented as a basic python HTTP server. The process involves several steps: 1) Receive a POST request with the URL of the data being requested and optionally an ID and asymmetrically encrypted phrase to verify the requester's identity. CageCoach implements this with RSA public/private key pairs. 2) Locate the data profile for the requested data on the DCS server, itself a simple text file containing pointers to decrypt and identify the requested data. Additionally, if the user is registered with the DCS (registry comprised of a private key for decrypting phrases, the plain text passphrase, and a id name) it will load their profile. We implement this as simply a separate json file containing each requester's information. Our approach assumes this registry happens outside of the CageCoach architecture but can utilize it. 3) The DCS will download the encrypted file from its host using HTTP GET. After reception, the DCS will decrypt the data file and load the redaction policies that match the specific data item (by its name), the data type (json, mp3, etc), and finally the policies for the requester (if provided). CageCoach implements this encryption with symmetric keys using pythons Fernet library. 4) The DCS will apply these redaction operations, gradually chipping away data until left with whatever is allowed to pass. 5) The remaining data is sent to the requester in the POST response. The specifics of how the redaction is applied is outlined in the next section.

\subsection{Access Control and Redaction Pipelines}
\begin{figure}
    \centering
    \includegraphics[width=0.48\textwidth, height=0.27\textwidth]{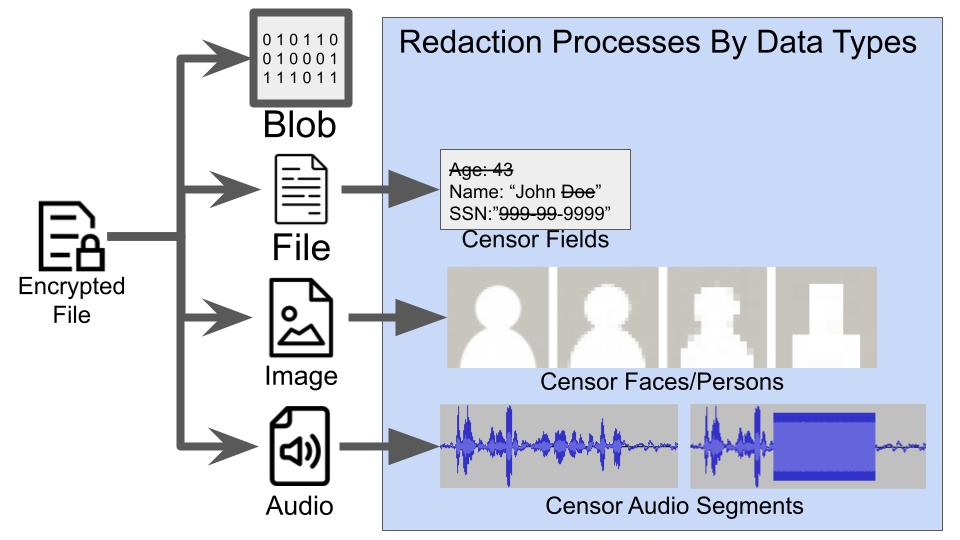}
    \caption{CageCoach Redaction Pipeline, providing a generalized measure of privacy assurance.}
    \label{fig:redaction-pipeline}
\end{figure}

\begin{figure}
    \centering
    \includegraphics[width=0.48\textwidth, height=0.27\textwidth]{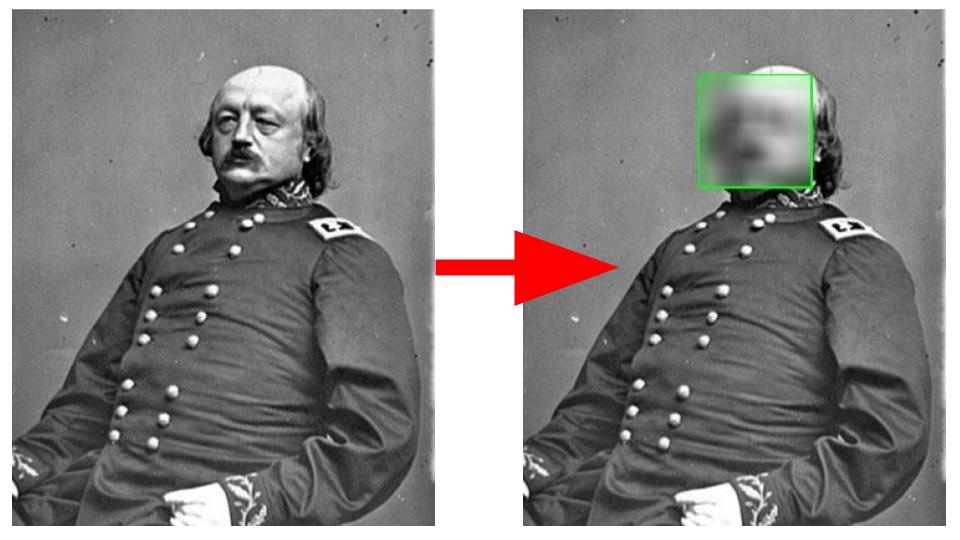}
    \caption{CageCoach Redaction Pipeline example blurring a specific image.}
    \label{fig:redaction-pipeline-example}
\end{figure}

\begin{figure}
    \centering
    \includegraphics[width=0.48\textwidth, height=0.27\textwidth]{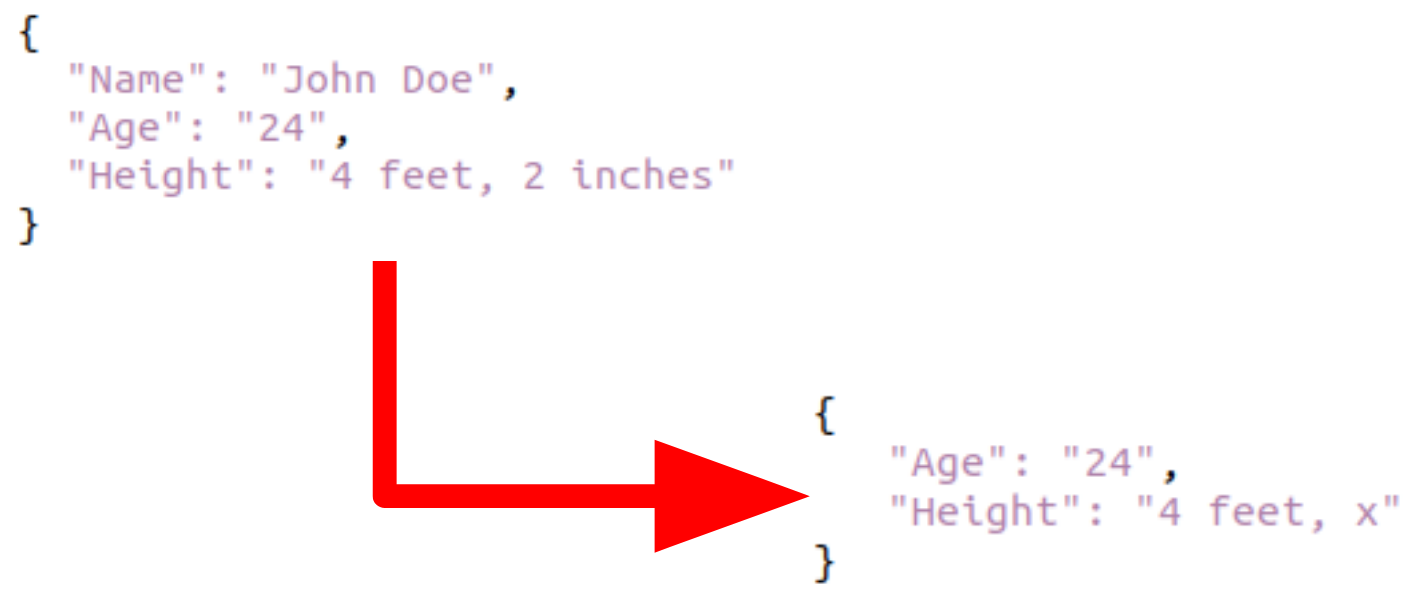}
    \caption{CageCoach Redaction Pipeline example redacting specific text and fields.}
    \label{fig:redaction-pipeline-example-2}
\end{figure}

CageCoach's read-only assumption for user data is not a binary, like older models of RBAC based system, but granular. By using a series of redaction operations over requested data, CageCoach can allow partial access to data. These operations, dividable by datatype as outlined in Fig. \ref{fig:redaction-pipeline}, provide for blurring faces in images, redacting text in jsons and csvs, and muting specific words or background noises recognized in audio. In the overall data request pipeline after a user has requested data and the DCS has downloaded the target data, it will apply these redaction operations according to the specific user, datatype, and data item. This provides three levels of granularity for controlling data flow outwards to requesters: by datatype (all jsons, csvs, mp3s, etc), by data item (ex: specific files like example-1.json hosted on Google Drive or example-2.json hosted on dropbox), and by requester id (ex: John Doe can access the user's name, but Jane Doe can only see the user's first name). However, such operations that would be specific to an owner, such as blurring only the owner's face, require the owner provide their own data to the redacting DCS. Our implementation we provide does general redaction such as blurring all faces and removing a handful of well known text fields such as social security and street addresses. We do not implement an audio redaction approach as there isn't a general python capable pre-built audio redaction library nor a common set of what "words" should be auto removed, unlike faces in images. CageCoach does support extensions to these operations to tailor to specific users. Our implementation uses the Haar cascade and OpenCV2 \cite{tools-cv2} python libraries for blurring faces (illustrated with the blurring of photo of American Union Army General Benjamin Butler fig. \ref{fig:redaction-pipeline-example}), and python Pandas to redact textual data (example of such in fig. \ref{fig:redaction-pipeline-example-2}).

\section{Conclusion}
In this work, we introduced a new sharing oriented implementation of DCFS: CageCoach. CageCoach streamlines the older RBAC heavy and trust-necessary hosting models of DFS, while using the simpler HTTP GET/POST ecosystem to facilitate easier data sharing. All of this is possible while still respecting the privacy of users through granular customizeable redaction pipelines that handle removal of sensitive user information.

\section{Limitations and Future Work}
CageCoach has a set of drawbacks and limitations. CageCoach is implemented as a demonstration of a new interpretation of sharing-oriented DCFS and not intended for industrial or commercial use. Future implementations would need to provide better integration with hosting services like Google and Dropbox, and provide tougher and more robust security checks and infrastructure. Likewise future work improvements would be needed to make the redaction operations more capable and workable on a wider set of diverse data. Notably there are two non-implementation limitations that stunt CageCoach and the broader goal of safe sharing oriented DFS:

\noindent{\textbf{$\bullet$ No system can stop external data reconstruction.}}

No matter if a user is using CageCoach, Google Drive, or any other hosting system, external actors with access to pieces of separate data can always reassemble it together. For example, an actor A has access to a subset of data 1, and an actor B has access to another subset of data 1. These two actors are not allowed access to either subset of data by the policies of the user whose data it is. However, this does not stop nor disincentivise actor A and B from simply sharing with each other the user's data. Each filling in the other's gap of missing data. No system can solve this if the requesting actors are able to observe data.

\noindent{\textbf{$\bullet$ Leakage is still possible through indirect implicating fields.}}

CageCoach's redaction pipeline is quite rudimentary, in some cases data may be leaked through a combination of unrelated fields. For example, with a street address, a malicious user may be able to correctly guess a zip code when paired with other information. This is due to CageCoach's inability to understand the connections between data.

CageCoach's unique sharing-oriented DCFS structure provides several new areas of exploration. CageCoach itself can be expanded to cover more datatypes, and work can be done to integrate the ingress of user's data to the data hosts that CageCoach manages.

\subsection{Collective Redaction Rules For Multi-Owner Data}
Given our system's usage of a redaction pipeline, one could envision a scenario where data that is collected by one user, but contains multiple other users' data is pass around each impacted user's DCS for specific group based redaction. This would facilitate greater granularity of redaction and a sense of group ownership over data and its privacy implications.

\subsection{Enhanced ACL And Redaction Through Impact Trees}
A future work could examine how to enhance the redaction rules to include field implications to provide greater coverage of privacy in the event a user misses these concepts themselves. This would fill in the gaps that leaking implicating fields create.

\bibliographystyle{ACM-Reference-Format}
\bibliography{refs.bib}

\appendix

\end{document}